# A high-efficiency neuroevolution potential for tobermorite and calcium silicate hydrate systems with ab initio accuracy

Xiao Xu[1], Shijie Wang[1], Haifeng Qin[1], Zhiqiang Zhao[2,3*], Zheyong Fan[4], Zhuhua Zhang[3*] and Hang Yin[1,5*]

1 *College of Water Conservancy and Civil Engineering, Shandong Agricultural University, Tai'an, 271018, China*

2 *Institute of High-Performance Computing (IHPC), Agency for Science, Technology and Research (A∗STAR), 1 Fusionopolis Way,#16-16 Connexis, Singapore, 138632, Singapore*

3 *State Key Laboratory of Mechanics and Control for Aerospace Structures, Key Laboratory for Intelligent Nano Materials and Devices of Ministry of Education, and Institute for Frontier Science, Nanjing University of Aeronautics and Astronautics, Nanjing 210016, China*

4 *College of Physical Science and Technology, Bohai University, Jinzhou 121013, P. R. China*

5 *School of Science, Harbin Institute of Technology, Shenzhen 518055, China*

\* Corresponding authors' Email address:

zhiqiangzhao55@163.com (Zhiqiang Zhao), chuwazhang@nuaa.edu.cn (Zhuhua Zhang) and yinh@sdau.edu.cn (Hang Yin)

**Abstract:**

Tobermorite and Calcium Silicate Hydrate (C-S-H) systems are indispensable cement materials but still lack a satisfactory interatomic potential with both high accuracy and high computational efficiency for better understanding their mechanical performance. Here, we develop a Neuroevolution Machine Learning Potential (NEP) with Ziegler-Biersack-Littmark hybrid framework for tobermorite and C-S-H systems, which conveys unprecedented efficiency in molecular dynamics simulations with substantially reduced training datasets. Our NEP model achieves prediction accuracy comparable to DFT calculations using just around 300 training structures, significantly fewer than other existing machine learning potentials trained for tobermorite. Critically, the GPU-accelerated NEP computations enable scalable simulations of large tobermorite systems, reaching several thousand atoms per GPU card with high efficiency. We demonstrate the NEP's versatility by accurately predicting mechanical properties, phonon density of states, and thermal conductivity of tobermorite. Furthermore, we extend the NEP application to large-scale simulations of amorphous C-S-H, highlighting its potential for comprehensive analysis of structural and mechanical behaviors under various realistic conditions.

# 1. Introduction

In modern cement materials science, atomistic investigations have become essential for elucidating fundamental atomic-scale mechanisms and enabling the rational design of advanced cement materials. Over the past decade, there has been a surge of interest in understanding the atomic and molecular interactions within cementitious phases, particularly Calcium Silicate Hydrate (C-S-H) － the principal hydration product of cement[1-6]. The nanoscale structure of C-S-H exhibits structural similarities to the crystalline tobermorite phase, and plays a critical role in determining the mechanical strength, durability, and macroscopic performance of concrete[7-9]. The studies of these nanoscale effects yields profound insights into the intrinsic behavior of cementitious systems, paving the way for next-generation construction materials with tailored properties and improved sustainability. Tobermorite features a layered structure consisting of calcium oxide sheets separated by silicate chains and water molecules. The structure is further complicated by multiple polymorphs including 9 Å, 11 Å and 14 Å tobermorite, distinguished by the interlayer spacing and variations in the silicate chain arrangements. A series of studies have been conducted to investigate the crystallographic and mechanical properties of tobermorite using experimental techniques[10], density functional theory (DFT) calculations[11, 12], and molecular dynamics (MD) simulations[13, 14]. These complementary approaches have collectively advanced the understanding of tobermorite's structural characteristics and mechanical performance, contributing to the broadened knowledge of C-S-H phases in cementitious materials.

For C-S-H, DFT-based *ab initio* methods serve as a powerful tool for obtaining benchmark values of key mechanical properties such as elastic constants and moduli[15-17]. These quantum mechanical methods further enable detailed investigations of the hydration formation process and provide atomic-level insights into the structural characteristics and water content of C-S-H [18, 19]. While *ab initio* molecular dynamics (AIMD) offers the advantage of simulating dynamic evolution processes without relying on empirical force field parameters, its applicability is constrained by system size limitations (typically several hundred atoms) and short simulation timescales (<1 ns), along with expensive computational cost[20]. To overcome these limitations, various classical force fields have been developed for C-S-H studies, each showing distinct advantages yet obvious limitations. The ClayFF[21], while compatible with organic force fields and a variety of aqueous ions, may not accurately reproduce certain properties of C-S-H like elasticity. The CSHFF force filed[22], a reparametrization of ClayFF, offers improved representation of C-S-H systems but necessitates charge neutralization adjustments that may introduce computational uncertainties. On the other hand, ReaxFF[23] stands out for its ability to model both disordered and ordered phases and to simulate chemical reactions within C-S-H, but its high computational cost and lack of specific parametrization for cement-related elements have limited broad applications. These classical force fields have been widely used to study structure-property relationships of C-S-H, particularly the influence of temperature and compositional variables (water content, Ca/Si ratio) on mechanical performance[17]. They have also been employed to investigate shear and creep behaviors of tobermorite and have shown potential in thermal conductivity studies[24-26]. Meanwhile, improved classical force fields such as CEMFF[27] and ERICAFF[28] have been successively proposed, incorporating more comprehensive parameterization schemes and broader performance validation across diverse properties. Despite these advancements, the selection of an appropriate force field for C-S-H systems remains challenging, as significant uncertainties persist regarding their transferability and accuracy across different simulation scenarios [29].

In other ways, the rise of machine learning potentials[30, 31] (MLPs) has provided a promising way for modeling atomic interactions with *ab initio* accuracy. For instance, Kobayashi *et al.*[32] used artificial neural networks (ANNs) to construct MLPs for tobermorite systems, using energy and force data as objective functions during training. Their results demonstrated that the ANN-based MLPs could accurately reproduce basic lattice constants, elastic constants, and vibrational density of states, achieving fidelity comparable to first-principles calculations. Zhou and his group[33-35] have also constructed MLPs for tobermorite and C-S-H using the deep potential (DP) method[36, 37]. Compared to empirical force fields, these MLPs significantly improved the accuracy of predicted structural characteristics and mechanical behavior in tobermorite. The DP model exhibited strong transferability, as validated by accurate predictions of C-S-H systems across a range of calcium-to-silicon (C/S) ratios. More recently, Zhu[38] constructed MLPs for 9 Å, 11 Å, and 14 Å tobermorite systems using the neural equivariant interatomic potential[39] (NequIP) framework, demonstrating the great potential of NequIP to support large-scale MD simulations of up to tens of thousands of atoms.

Despite these advances, most existing MLP models rely on extensive DFT-labeled datasets and remain challenging to scale effectively for realistic cementitious simulation systems involving millions of atoms. Moreover, the lack of publicly available models hinders reproducibility and broader evaluation. Notably, Fan *et al.*[40] developed the neuroevolution potential (NEP) framework [41, 42] based on GPUMD software, which greatly reduces the dependence on the number of training sets by optimizing the model training architecture. Meanwhile, the computational efficiency of NEP shows advantages over other related MLP models while maintaining comparable computational accuracy. The GPU-accelerated architecture of NEP model further enhances computational performance, making it particularly advantageous for running large-scale simulations. The NEP model has been successfully applied to a variety of material systems, such as silicon[43], carbon[44, 45], water[46], and complex alloy systems[47, 48], confirming its capability to accurately capture interatomic interactions under a wide range of temperature and pressure conditions. These characteristics render NEP a highly promising framework for developing reliable atomistic interaction models in complex engineering materials such as tobermorite and C-S-H.

To address these issues, we adopt the NEP framework, optimized via an active learning strategy that significantly reduces training datasets while maintaining high accuracy. The NEP approach exhibits superior scalability and efficiency, particularly when harnessing GPU computing capabilities. This study not only establishes NEP's high accuracy and computational performance but also explores its applicability in predicting thermal conductivity and mechanical properties, extending to amorphous C-S-H systems representative of practical cementitious materials.

## 2. Comparison of machine learning potentials

### 2.1 Neuroevolution potential（NEP）

We here use the NEP framework to develop a MLP for the tobermorite system. The NEP framework use a feedforward neural network (FNN) to represent the atomic site energy as a function of a descriptor components. The mapping from the input layer to the output layer is generally a nonlinear function, and its specific form depends on the number of hidden layers in the neural network. In the current implementation, the NEP framework uses only a single hidden layer, enabling the site energy of atom $i$, $U_i$, to be expressed as a composite function:

$$U_i = \sum_{\mu=1}^{N_{neu}} w_\mu^{(1)} \tanh\left(\sum_{\nu=1}^{N_{des}} w_{\mu\nu}^{(0)} q_\nu^i - b_\mu^{(0)}\right) - b^{(1)}, \quad \#1)$$

where tanh($x$) is the activation function for the hidden layer，$N_{des}$ is the dimension of the input layer (descriptors components), while $N_{neu}$ is the dimension of the hidden layer, i.e., the number of neurons in the hidden layer.

In the NEP model [40], the radial functions in the atomic environment descriptors were optimized by improving the combination of basis functions, and angle descriptors with higher-order correlations were added, which can clearly reflect the local environment of atomic interactions. For a given central atom *i*, a set of radial descriptors ($n \geq 0$) is defined as follows:

$$q_n^i = \sum_{j \neq i} g_n(r_{ij}), \#(2)$$

where $g_n(r_{ij})$ is a radial function, which is controlled solely by the radial cutoff distance. It is defined as a linear combination of $N_{bas}^R + 1$ basis functions $\{f_k(r_{ij})\}_{k=0}^{N_{bas}^R}$.

In the angular descriptor components, the angular function $g_n(r_{ij})$ takes a form similar to that of the radial descriptors. For the angular descriptors, expressions for both three-body and four-body descriptors are considered here. For the three-body descriptor $q_{nl}^i$ ($0 \leq n \leq n_{max}^A$, $1 \leq l \leq l_{max}^{3b}$) and the four-body descriptor $q_{nl_1l_2l_3}^i$ ($0 \leq n \leq n_{max}^A$, $1 \leq l_1 \leq l_2 \leq l_3 \leq l_{max}^{4b}$), they are defined as follows:

$$q_{nl}^i = \sum_m (-1)^m A_{nlm}^i A_{nl(-m)}^i, \#(3)$$

$$q_{nl_1l_2l_3}^i = \sum_{m_1=-l_1}^{l_1} \sum_{m_2=-l_2}^{l_2} \sum_{m_3=-l_3}^{l_3} \binom{l_1}{m_1 m_2 m_3} A_{nl_1m_1}^i A_{nl_2m_2}^i A_{nl_3m_3}^i, \#(4)$$

$$A_{nlm}^i = \sum_{j \neq i} g_n(r_{ij}) Y_{lm}(\theta_{ij}, \phi_{ij}), \#(5)$$

where $Y_{lm}(\theta_{ij}, \phi_{ij})$ is the spherical harmonic function of the polar angle $\theta_{ij}$ and the azimuthal angle $\phi_{ij}$ from atom *i* to atom *j*. For the training of NEP model, the total loss function is composed of a weighted sum of several components:

$$L(z) = \lambda_e L_e(z) + \lambda_f L_f(z) + \lambda_v L_v(z) + \lambda_1 L_1(z) + \lambda_2 L_2(z), \#(6)$$

where $L_e(z)$, $L_f(z)$ and $L_v(z)$ represent the root mean square errors (RMSE) between the reference and predicted values for energy, force, and virial, respectively. Additionally, regularization loss functions are considered, denoted as $L_1(z)$ and $L_2(z)$.

## 2.2 DP and NequIP

The DP method employs ANNs to accurately represent potential energy surfaces (PESs) based on data from DFT calculations. Feedforward neural networks are used to automatically construct descriptors to describe the local environment of each atom. Data starts at the input layer, is processed by nodes in multiple hidden layers, and is finally passed to the output layer. This method builds an ANN for each atom, for a given atom $i$, and assigns it a local environment containing a set of neighboring atoms around it. The local coordinate descriptor $\{D_{ij}\}$ of the corresponding atom is described by defining the coordinates of all atoms within the truncated radius.

$$D_{ij} = \left\{\frac{1}{R_{ij}}, \frac{x_{ij}}{R_{ij}^2}, \frac{y_{ij}}{R_{ij}^2}, \frac{z_{ij}}{R_{ij}^2}\right\}, \#(7)$$

where $D_{ij}$ is the geometric position between the atoms. $R_{ij}$ denotes the vector of the atom $i$ pointing to the atom $j$, and $(x, y, z)$ is the Cartesian component of the vector $R_{ij}$ in the local coordinate system.

The NequIP framework is built upon an equivariant graph neural network (GNN) architecture, which eliminates the need for manually crafted descriptors. It is capable of constructing accurate interatomic potentials from a limited dataset of fewer than 1000, or even as few as 100 DFT labeled structures. The model learns representations on the atomic graph from invariant geometric features such as radial distances or angles. Unlike traditional FNNs, GNNs incorporate symmetry constraints. While using relative position vectors, they also introduce features that include both scalars and higher-order geometric tensors. These tensors are iteratively computed through a series of convolutions over the distances to neighboring atoms. The convolution operation combines the product of neighboring features with the tensor product of a radial function R(r) and the spherical harmonic projection of the unit vector $\hat{r}_{ij}$, as follows:

$$S_m^{(l)}(r_{ij}) = R(r_{ij})Y_m^{(l)}(\hat{r}_{ij}), \#(8)$$

If $\hat{r}_{ij}$ denotes the relative position vector from the central atom $i$ to a neighboring atom $j$, then $\hat{r}_{ij}$ and $r_{ij}$ represent the associated unit vector and the interatomic distance, respectively. $R(r_{ij})$ refers to the rotation-invariant radial function, which outputs the radial functions that govern the interactions between all filter-feature vector pairs.

Compared with DP and NequIP, NEP employs a combination of radial basis functions and neural network. The FNN used in NEP does not enforce data symmetry, and its output depends directly on the input. With only a single hidden layer, the model is structurally simpler and more flexible. This simplicity makes NEP more effective in training stable and reliable potential functions, especially when the available dataset is limited. This advantage primarily stems from NEP's use of local descriptors integrated with physical features, which enables more efficient learning of the potential energy surface, reduces the reliance on large quantities of high-precision data, and enhances both training efficiency and model stability. Compared with DP and NequIP, NEP adopts a combination of radial basis function and neural network, and the FNN used does not consider the symmetry of the data, and the output directly depends on the input, and only introduces a hidden layer, which is simpler in structure and more flexible.

## 3. Modeling and estimation of NEP

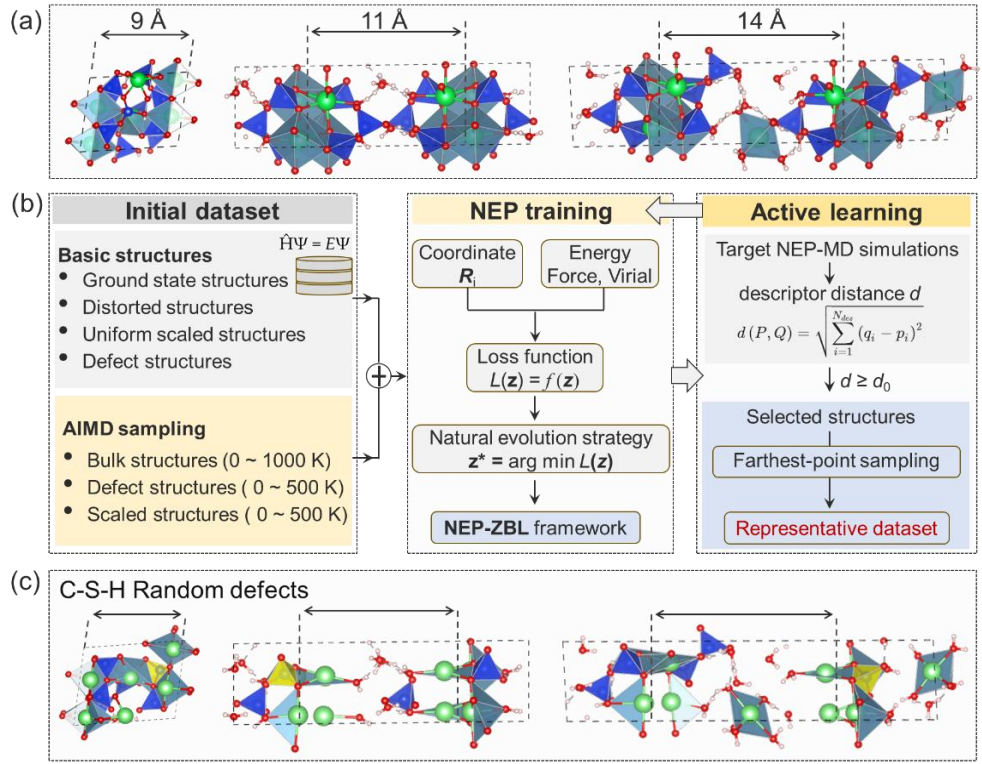

**Figure 1.** (a) Atomic structures of tobermorite 9 Å, 11 Å, and 14 Å. (b) The workflow for NEP model training. (c) Random defect structure of C-S-H (yellow regions indicate the distribution of randomly located silicate tetrahedra).

Due to their nearly identical microscopic structure and chemical compositions, tobermorite is widely regarded as archetypal model for constructing C-S-H[7] phases. To ensure consistency with previous work and facilitate direct comparison, we adopted the same atomistic tobermorite models from studies[32, 33, 38] for training dataset generation and NEP model development. As shown in Figure 1(a), the silicon tetrahedra in tobermorite are arranged according to the Dreierketten rule, forming a layered structure and interleaved calcium layers. Depending on the degree of hydration, tobermorite can be classified into three variants based on their interlayer distances, namely 9 Å, 11 Å, and 14 Å. Notably, the 11 Å and 14 Å variants contain significantly more interlayer water than the 9 Å variant.

High-quality training datasets are prerequisite for constructing high-precision machine-learning force fields. In this work, all atomic configuration in dataset were labeled by DFT calculations using the VASP code[49]. Electronic interactions were described with the Perdew-Burke-Ernzerhof (PBE) exchange-correlation functional within the generalized gradient approximation (GGA), in conjunction with the projector augmented wave (PAW) method. The cutoff energy for the plane-wave basis set was set to 600 eV. The convergence criteria for energy and forces were set to $10^{-5}$ eV and 0.03 eV/Å, respectively. To improve accuracy, the DFT-D3 van der Waals (vdW) dispersion corrections were also included to capture long-range dispersion interactions. The Γ-centered $k$-point mesh was generated using the Monkhorst-Pack scheme, with the number of points determined by the formula $\max(1, |\mathbf{G}_i|/\Delta k)$, where $\mathbf{G}_i$ ($i = a,b,c$) are the reciprocal lattice vectors. For AIMD simulations, $\Delta k$ was set to 0.6 Å$^{-1}$, and for static calculations, it was set to 0.25 Å$^{-1}$. All structural manipulations and analysis of the simulation results were performed using the ASE[50] and PyNEP[40] libraries in Python.

As illustrated in Figure 1(b), the NEP training workflow comprises three main steps: initial training set construction, NEP model training, and active learning sampling. To construct the initial training set, we perform geometric optimization on the initial structures of tobermorite 9 Å, 11 Å, and 14 Å. These optimizations were limited to a maximum of 50 ionic steps to efficiently obtain the ground-state configurations. To enhance structural diversity, each optimized structure was subjected to ±10% volume scaling, combined with random perturbations involving box deformations of 1.5% and atomic displacements of 0.05 Å. Subsequently, AIMD simulations are performed under the *NVT* ensemble for these perturbed structures at temperatures of 100 K, 300 K, 800 K, and 1500 K, using a time step of 2.0 fs. During the AIMD simulations, snapshots were extracted from the trajectories every 300 fs. Additionally, to evaluate the potential of NEP for modeling C-S-H, we constructed defect structures with higher Ca/Si ratios by randomly removing silicate chains at different positions, as shown in Figure 1(c).

Through the above steps, the initial training dataset, encompassing configurations with a range of volumes, defects, and strain states, was constructed. All structures in the training dataset were subsequently subjected to static calculations to obtain accurate atomic positions, energies, forces, and virials for use in NEP model training. The NEP training was then performed using the GPUMD package, with the Separable Natural Evolution Strategy (SNES) employed to optimize the hyperparameters by minimizing the loss function. To improve the robustness and training efficiency of the MLP, we use a hybrid framework that integrates the empirical short-range repulsive Ziegler-Biersack-Littmark (ZBL) potential[51] into the NEP model. This hybrid NEP-ZBL approach significantly enhances the robustness of developed MLP model by mitigating unphysical atomic clustering and preventing catastrophic failures during long-time MD simulations. An active learning strategy is introduced to iteratively sample structures from a serial of NEP-MD trajectories using the farthest point sampling method. These NEP-MD trajectories cover a wide range of thermodynamic variables (temperature and pressure) and deformation modes (tension, compression and shear). We calculated the descriptor distance for each snapshot and then selected the most structurally diverse structures from targeted NEP-MD trajectories. These newly selected configurations have relatively large descriptor distances to existing configurations in the descriptor space and are then computed using DFT to serve as additional training datasets. By combining the initial dataset with the newly added configurations, the next-generation NEP model is retrained, and the above simulation and selection steps are repeated. This iterative active learning approach allows the NEP model to more comprehensively cover the atomic configuration space encountered in MD simulations, thereby ensuring higher accuracy and better transferability. The final training set consists of a total of 302 structures: 120 for tobermorite 9 Å, 74 for 11 Å, and 108 for 14 Å, respectively. Compared to existing MLPs, this represents a reduction in the number of training structures by 1-2 orders of magnitude. Detailed values are shown in Table 1.

**Table 1.** The number of structures in the training and test datasets for NEP and other MLPs [32, 33, 38].

|        | Training set | Test set |
|--------|--------------|----------|
| MLP-FE | 35331        | 3926     |
| MLP-E  | 70663        | 7851     |
| DP     | 6300         | 700      |
| NequIP | 1425         | 8000     |
| NEP    | 302          | 40       |

We here set the cutoff radius for both the radial and angular descriptors to 4.5 Å. The radial and angular descriptors were expanded using Chebyshev polynomials, with the radial expansion order set to 10 and the angular expansion order set to 8. The descriptor functions were constructed using 10 and 8 basis functions, respectively. The neural network contained 50 neurons in each hidden layer. The weight factors in the loss function were set as follows: $\lambda_1 = \lambda_2 = 0.05$, $\lambda_e = \lambda_f = 1.0$, and $\lambda_v = 0.2$. The training was performed using the SNES algorithm with a batch size of 500 structures over a total of one million generations. The detailed hyperparameter settings are listed in Table 2. The final NEP models fitted in this study will be released in an open repository[52].

**Table 2.** NEP Training Hyperparameter Settings.

| Hyperparameter | Data |
| --- | --- |
| version | 4 |
| type | 4 Ca H O Si |
| cutoff | 4.5 4.5 |
| n_max | 10 8 |
| basis_size | 10 8 |
| l_max | 4 2 |
| neuron | 50 |
| lambda_1 | 0.05 |
| lambda_2 | 0.05 |
| lambda_e | 1 |
| lambda_f | 1 |
| lambda_v | 0.2 |
| zbl | 2 |
| batch | 500 |
| population | 50 |
| generation | 1000000 |

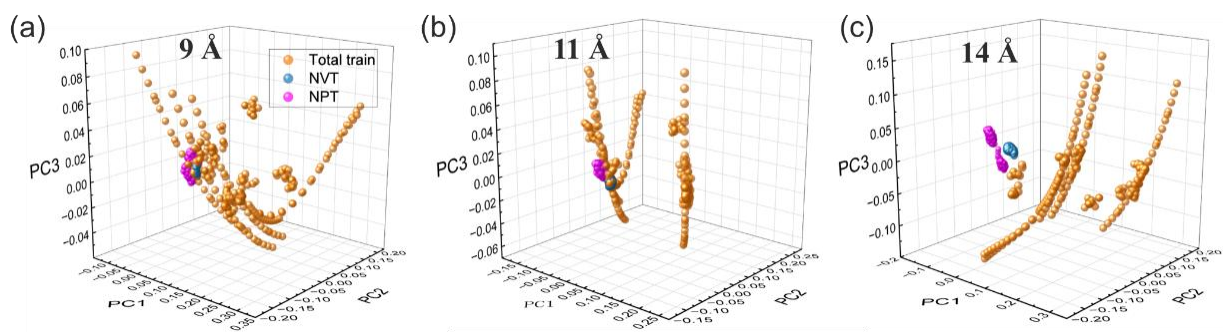

**Figure 2.** Distribution of structures in the three-dimensional PC subspace. All structures are from training dataset and extracted from NVT and NPT MD simulations at 300 K.

To assess the reliability of the trained NEP model in MD simulations at finite temperatures, we compare the distribution in the descriptor space for structures from both *NVT* and *NPT* MD simulations with the training structures. The high-dimensional descriptor space can be effectively reduced by principal component (PC) analysis[53]. Figure 2 presents the three-dimensional distribution of the sampled configurations from MD trajectories within the PC space. It is evident that the NEP model accurately captures the configurational space explored during the MD simulations for the 9 Å and 11 Å, with the corresponding data points well covered within the training dataset. In contrast, the configurations associated with the 14 Å structure are more widely dispersed in the PC space. This can be attributed to the intrinsic complexity of the 14 Å structure [54], where the larger interlayer spacing and the more diverse arrangement of hydrated ions introduce a broader range of structural states during MD sampling. Many of these thermodynamic structures are not adequately represented in the original training dataset, thereby reducing the effectiveness of the PC analysis in achieving dimensionality reduction for this system.

## 4. Results and discussion

### 4.1 Accuracy and Computational Efficiency of the NEP Model

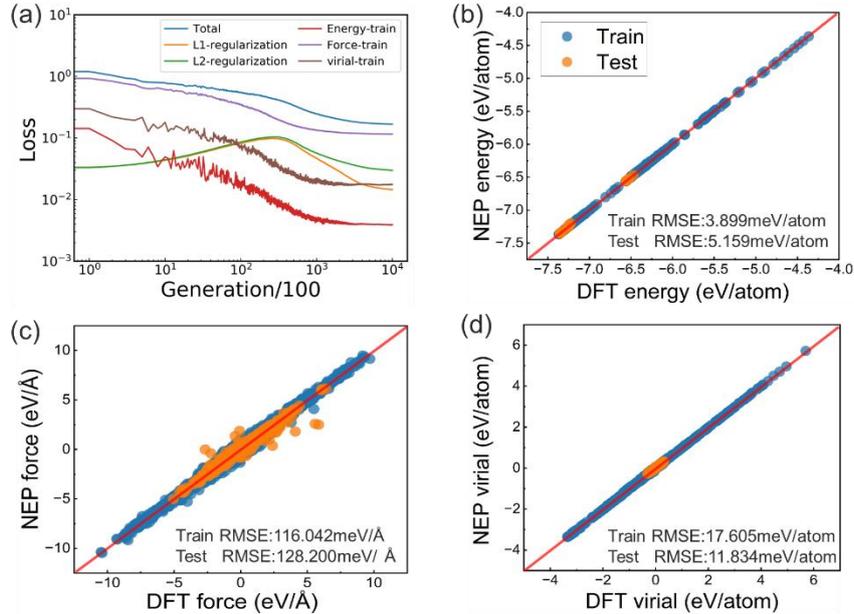

**Figure 3.** Performance of the NEP model. (a) Loss function as a function of training steps, (b) Comparison of energies, (c) Forces, and (d) Virial stresses between NEP predictions and DFT calculations.

After the training was completed, we obtained a highly accurate NEP model. To verify the generalization ability of the developd NEP model, we further generated a test set, with a training-to-test set ratio of 8:1. Figure 3(a) illustrates the evolution of the loss function for different components as a function of training steps. The loss function exhibits a clear convergence trend after approximately one million training steps, indicating that the training of the NEP model has reached optimal performance. To further validate the accuracy of the NEP model, we compared the predicted energies, forces, and virial stresses [Figure 3(b-d)] against the corresponding reference values obtained from DFT calculations. The results show that all data points are closely clustered around the $y = x$ reference line, demonstrating that the NEP model achieves high fitting accuracy for the energies, forces, and virial stresses across all configurations in the training set. Furthermore, the model exhibits excellent predictive performance on the test set, with RMSE of 5.159 meV/atom for energy, 128.200 meV/Å for force, and 11.834 meV/atom for virial. These results indicate that the trained NEP model can accurately predict the energies and forces of the tobermorite system, achieving a level of accuracy comparable to that of DFT calculations and thereby demonstrating outstanding overall performance.

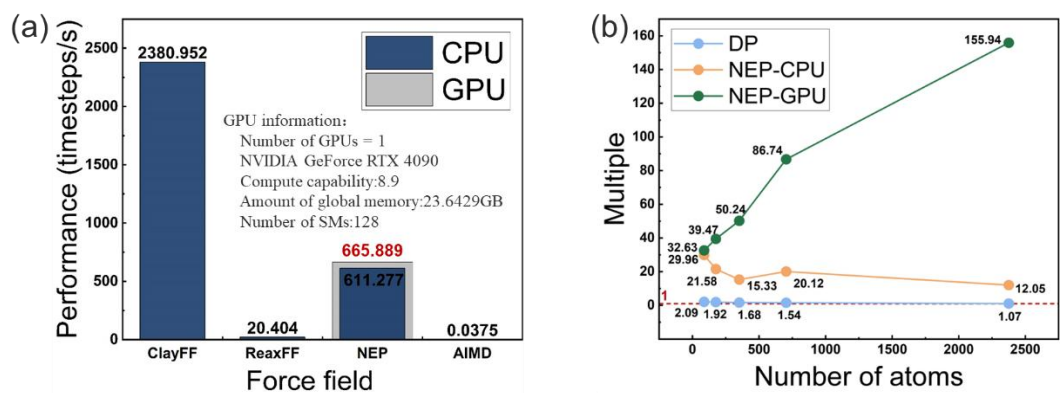

**Figure 4.** (a) Comparison of computational speeds for different potential models (LAMMPS with 12 CPU cores), (b) Computational speed comparison between DP and NEP (the red dashed line represents ReaxFF as the reference, set to 1, the *y*-axis indicates the speed-up factor).

To evaluate the computational efficiency of the NEP model, *NVT* simulations are performed at 300 K for a total duration of 2 ns using the initial 11 Å structure comprising 88 atoms. The results are presented in Figure 4(a). Although the NEP model is several times slower than the classical ClayFF force field when implemented in the LAMMPS-CPU package, it significantly outperforms ReaxFF, achieving a speed advantage exceeding one order of magnitude. Moreover, the NEP model achieves a computational speed-up of approximately four orders of magnitude compared to AIMD, highlighting its excellent balance between accuracy and efficiency. When running on a single NVIDIA GeForce RTX 4090 GPU, the NEP model achieves a computational speed of 665.889 timesteps per second, representing a substantial improvement over its CPU performance. A comparative analysis with the DP model, as reported in the literature[33], is shown in Figure 4(b). Across various system sizes, NEP consistently outperforms DP in computational speed. For small systems, the NEP model can be up to ten times faster than DP. Even for systems exceeding 2,250 atoms, NEP maintains approximately a tenfold speed advantage. Notably, the GPU-accelerated NEP model exhibits increasing computational speed with larger system sizes. In systems comprising thousands of atoms, NEP can outperform DP by several hundred times. These results highlight the significant benefits of GPU acceleration, which markedly enhances the scalability and efficiency of NEP. While the speed-up is less dramatic for small-scale systems, NEP-GPU consistently outperforms both DP and NEP-CPU implementations in large-scale MD simulations, demonstrating its strong potential for efficient modeling of complex materials systems.

## 4.2 Comparison of NEP with DFT and Other MLPs

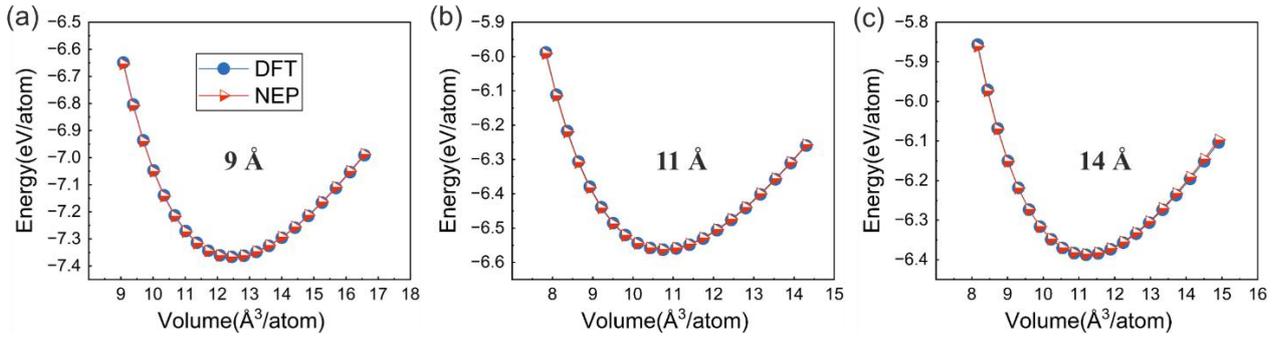

**Figure 5.** EOS curves of tobermorite calculated by DFT and NEP for (a) 9 Å, (b) 11 Å, and (c) 14 Å.

Figure 5 presents the equation of state (EOS) curves for the three tobermorite structures as calculated by DFT and the NEP model. For each point on the EOS curve, the initial supercell was uniformly scaled to achieve a target volume (corresponding to ±10% volumetric strain range). Subsequently, both the cell shape and atomic positions were relaxed to equilibrium using DFT calculations while keeping the volume fixed. To ensure a consistent comparison, the energies and volumes for the same set of equilibrium structure were evaluated using both NEP and DFT. As shown in the Figure 5, for all structures, the energy per atom predicted by the NEP model closely matches the corresponding DFT values. Across the entire tested volume range, the RMSE between NEP and DFT remains below 1 meV/atom, demonstrating the high fidelity of the NEP predictions. The ability of the NEP model to accurately reproduce the EOS is critical for reliably predicting the mechanical behavior of the material under various volume strains and pressure conditions. Table 3 summarizes the equilibrium lattice constants of tobermorite as predicted by DFT, DP, NequIP, and the NEP model. The NEP-predicted lattice constants show close agreement with those obtained from DP and NequIP, indicating comparable levels of accuracy among these MLPs in describing basic structure properties. Moreover, the maximum deviation of NEP-predicted lattice constants from the experimental values is within 1.4%, and the maximum deviation from the DFT results is within 0.8%, further confirming the high predictive accuracy and reliability of the NEP model.

**Table 3.** Lattice constants of tobermorite 9 Å, 11 Å, and 14 Å obtained by DFT and different MLPs. The percentage errors relative to experimental values [54-56] are shown in parentheses.

| Type | Parameters | Exp. | DFT | DP | NequIP | NEP |
|---|---|---|---|---|---|---|
| 9 Å | $a$ (Å) | 11.16 | 11.16(0.00) | 11.24(0.72) | 11.23(0.63) | 11.17(0.09) |
| | $b$ (Å) | 7.30 | 7.35(0.68) | 7.40(1.37) | 7.32(0.27) | 7.34(0.55) |
| | $c$ (Å) | 9.57 | 9.6(0.31) | 9.58(0.10) | 9.72(1.57) | 9.61(0.42) |
| | $\alpha$ (°) | 101.08 | 101.28(0.20) | 101.01(0.07) | 101.58(0.49) | 101.21(0.13) |
| | $\beta$ (°) | 92.83 | 92.76(0.08) | 90.81(2.18) | 90.92(2.06) | 92.89(0.06) |
| | $\gamma$ (°) | 89.98 | 90.27(0.32) | 89.97(0.01) | 91.34(1.51) | 90.21(0.26) |
| 11 Å | $a$ (Å) | 6.74 | 6.76(0.30) | 6.81(1.04) | 6.85(1.63) | 6.71(0.45) |
| | $b$ (Å) | 7.39 | 7.42(0.41) | 7.46(0.95) | 7.49(1.35) | 7.36(0.41) |
| | $c$ (Å) | 22.49 | 22.50(0.04) | 22.45(0.18) | 22.57(0.36) | 22.60(0.49) |
| | $\alpha$ (°) | 90.00 | 90.26(0.29) | 90.71(0.79) | 90.99(1.10) | 90.13(0.14) |
| | $\beta$ (°) | 90.00 | 90.64(0.71) | 90.02(0.02) | 89.25(0.83) | 90.80(0.89) |
| | $\gamma$ (°) | 123.25 | 123.03(0.18) | 123.52(0.22) | 123.79(0.44) | 122.79(0.37) |
| 14 Å | $a$ (Å) | 6.74 | 6.77(0.45) | 6.75(0.15) | 6.81(1.04) | 6.78(0.59) |
| | $b$ (Å) | 7.43 | 7.36(0.94) | 7.40(0.40) | 7.43(0.00) | 7.33(1.35) |
| | $c$ (Å) | 27.99 | 28.06(0.25) | 27.98(0.04) | 28.41(1.50) | 28.26(0.96) |
| | $\alpha$ (°) | 90.00 | 90.04(0.04) | 89.36(0.71) | 88.34(1.84) | 90.23(0.26) |
| | $\beta$ (°) | 90.00 | 89.06(1.04) | 88.57(1.59) | 88.91(1.21) | 89.23(0.86) |
| | $\gamma$ (°) | 123.25 | 123.99(0.60) | 123.38 (0.11) | 125.29(1.66) | 123.62(0.30) |

The elastic constants of tobermorite 9 Å, 11 Å, and 14 Å are calculated using the strain-energy method. In this approach, In this approach, a series of small deformations are introduced by applying incremental strains within elastic regime to the equilibrium structures. The force convergence criterion during structure relaxation is set to 0.0001 eV/Å. The elastic constants are derived by calculating the change in the total energy of the system before and after applying strain. According to elasticity theory, the total energy of the system is related to the strain as follows:

$$E_\varepsilon = E_0 + \frac{1}{2}C_{ij}\varepsilon_i\varepsilon_j, \#(10)$$

where $E_\varepsilon$ is the total energy after applying strain, $E_0$ is the total energy before applying strain, $\varepsilon_i$ and $\varepsilon_j$ are the strain components.

**Table 4.** Calculated elastic constants $C_{ij}$ of tobermorite 9 Å, 11 Å, and 14 Å obtained using different MLPs. The percentage errors relative to DFT results are shown in parentheses.

| $C_{ij}$ (GPa) | 9 Å | | | | 11 Å | | | | 14 Å | | | |
|---|---|---|---|---|---|---|---|---|---|---|---|---|
| | DFT | DP | NequP | NEP | DFT | DP | NequP | NEP | DFT | DP | NequP | NEP |
| $C_{11}$ | 175.20 | 167.74 (4.26) | 154.68 (11.71) | 173.59 (0.92) | 130.96 | 124.23 (5.14) | 117.52 (10.26) | 110.86 (15.35) | 102.33 | 77.10 (24.66) | 92.74 (9.37) | 96.20 (5.99) |
| $C_{12}$ | 58.79 | 63.61 (8.20) | 50.57 (13.98) | 52.19 (11.23) | 49.31 | 48.75 (1.14) | 46.14 (6.43) | 49.83 (1.05) | 39.28 | 43.49 (10.72) | 46.33 (17.95) | 40.81 (3.90) |
| $C_{13}$ | 36.94 | 37.39 (1.22) | 35.35 (4.30) | 34.42 (6.82) | 32.23 | 35.12 (8.97) | 31.75 (1.49) | 29.47 (8.56) | 20.75 | 18.44 (11.13) | 26.98 (30.02) | 24.58 (18.46) |
| $C_{22}$ | 167.96 | 160.10 (4.68) | 167.00 (0.57) | 150.40 (10.45) | 144.57 | 131.34 (9.15) | 112.83 (21.95) | 123.05 (14.89) | 97.63 | 81.58 (16.44) | 99.69 (2.11) | 96.57 (1.09) |
| $C_{23}$ | 38.54 | 34.72 (9.91) | 44.38 (15.15) | 41.23 (6.98) | 40.67 | 47.41 (16.57) | 37.22 (8.48) | 39.74 (2.29) | 34.07 | 22.91 (32.76) | 21.58 (36.66) | 21.05 (38.22) |
| $C_{33}$ | 104.72 | 100.27 (4.25) | 101.40 (3.17) | 115.48 (10.28) | 151.11 | 137.11 (9.26) | 126.81 (16.08) | 142.65 (5.60) | 48.83 | 75.72 (55.07) | 58.13 (19.05) | 59.89 (22.65) |
| $C_{44}$ | 50.64 | 36.48 (27.96) | 43.76 (13.59) | 43.57 (13.96) | 43.39 | 28.64 (33.99) | 30.90 (28.79) | 32.09 (26.04) | 38.43 | 24.05 (37.42) | 24.65 (35.86) | 26.69 (30.55) |
| $C_{55}$ | 12.86 | 5.86 (54.43) | 26.65 (107.23) | 6.19 (51.87) | 29.42 | 19.70 (33.04) | 19.21 (34.70) | 22.21 (24.51) | 16.64 | 16.73 (0.54) | 16.08 (3.37) | 20.23 (21.57) |
| $C_{66}$ | 39.74 | 46.44 (16.86) | 51.76 (30.25) | 48.26 (21.44) | 20.64 | 47.48 (130.04) | 42.87 (107.7) | 43.90 (112.69) | 19.45 | 31.59 (62.42) | 36.55 (87.92) | 40.40 (107.71) |

The elastic constants $C_{ij}$ predicted by the NEP model are systematically compared with those obtained from DFT and other MLPs, with the percentage errors relative to DFT reported for selected components in Table 4. Among all elastic constants, $C_{11}$, $C_{22}$, and $C_{33}$ are particularly important, as they directly reflect the stiffness and mechanical stability of the structure along different crystallographic directions and serve as key indicators of its structural properties. As shown in Table 4, the anisotropic trends in elastic constants computed by different MLPs are consistent, exhibiting only minor numerical deviations. Notably, for the principal diagonal components ($C_{11}$, $C_{22}$, and $C_{33}$), the NEP predictions are generally in closest agreement with the DFT results, showing the lowest percentage errors among all evaluated models.

Furthermore, the $C_{33}$ for tobermorite with interlayer spacings 9 Å and 14 Å are significantly lower than the corresponding $C_{11}$ and $C_{22}$, indicating strong anisotropy. This behavior is attributed to the layered structure of the 9 Å and 14 Å, which consist of alternating SiO$_4$ chains and calcium layers. The silicon chains are aligned along the $a$ and $b$ directions, while interlayer interactions are primarily governed by the spacing and ionic bonding between layers, leading to weaker mechanical response along the *c*-axis. In this direction, the interactions are mainly driven by weaker van der Waals forces and hydrogen bonds, resulting in the lowest mechanical strength along the *c*-axis. In contrast, tobermorite 11 Å structure exhibits the highest $C_{33}$ value, as the $SiO_4$ tetrahedra in the structure form a silicate network through Si-O-Si covalent bonds, which significantly enhances interlayer interactions and thus improves stiffness in the *c*-direction. Figure 6 further illustrates the deviations of the full elastic constants from the DFT results. The mean absolute errors of the stiffness matrices for the 9 Å, 11 Å, and 14 Å structures are 4.96 GPa, 8.97 GPa, and 7.53 GPa, respectively, indicating that NEP delivers highly accurate predictions of elastic properties. Overall, the NEP model demonstrates high accuracy in predicting the elastic constants of tobermorite systems, with percentage errors comparable to or even lower than those of DP and NequIP, suggesting that NEP holds clear advantages for simulating the mechanical properties of materials.

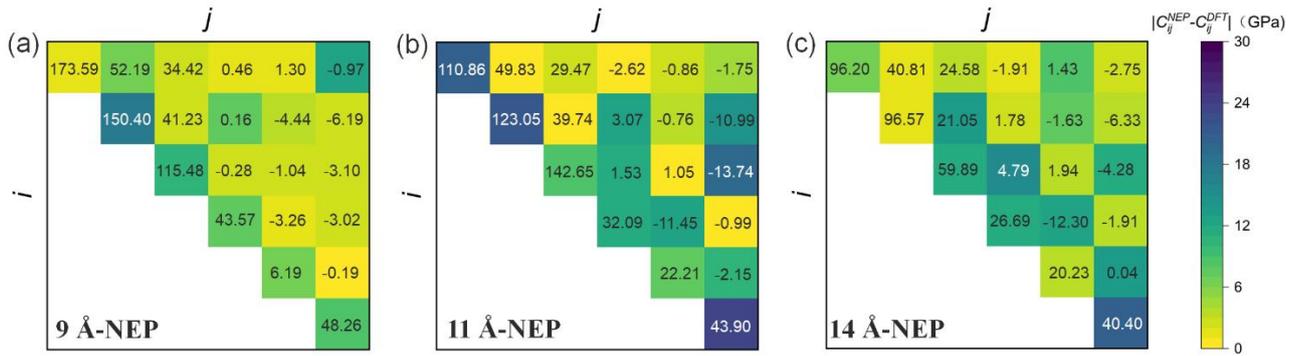

**Figure 6.** The elastic constants matrices of tobermorite (a) 9 Å, (b) 11 Å, and (c) 14 Å. The color represents the deviation between NEP and DFT.

**Table 5.** VRH approximation of bulk modulus, shear modulus, Young's modulus (GPa), and poisson's ratio of tobermorite 9 Å, 11 Å and 14 Å.

| Elastic constant | 9 Å | | | | 11 Å | | | | 14 Å | | | |
|---|---|---|---|---|---|---|---|---|---|---|---|---|
| | DFT | DP | NequIP | NEP | DFT | DP | NequIP | NEP | DFT | DP | NequIP | NEP |
| $K_V$(GPa) | 79.60 | 77.73 | 75.97 | 77.24 | 72.61 | 72.80 | 65.26 | 68.29 | 48.56 | 44.90 | 48.92 | 47.28 |
| $K_R$(GPa) | 71.04 | 56.99 | 70.25 | 73.80 | 69.73 | 71.43 | 62.57 | 66.24 | 37.24 | 43.27 | 42.60 | 42.22 |
| $K_{VRH}$(GPa) | 75.32 | 67.36 | 73.11 | 75.52 | 71.17 | 72.12 | 63.92 | 67.27 | 42.90 | 44.08 | 45.76 | 44.75 |
| $G_V$(GPa) | 41.55 | 37.25 | 43.95 | 40.38 | 35.70 | 36.59 | 34.73 | 36.81 | 25.22 | 24.44 | 25.83 | 28.55 |
| $G_R$(GPa) | 28.95 | 13.73 | 39.42 | 19.29 | 27.42 | 24.77 | 27.84 | 29.70 | 10.58 | 19.05 | 20.98 | 21.99 |
| $G_{VRH}$(GPa) | 35.25 | 25.49 | 41.68 | 29.84 | 31.56 | 30.68 | 31.29 | 33.25 | 17.90 | 21.75 | 23.40 | 25.27 |
| $E_{VRH}$(GPa) | 91.48 | 67.91 | 105.08 | 79.09 | 82.48 | 80.61 | 80.69 | 85.65 | 47.14 | 56.02 | 59.99 | 63.79 |
| $\mu$ | 0.30 | 0.33 | 0.26 | 0.33 | 0.31 | 0.31 | 0.29 | 0.29 | 0.32 | 0.29 | 0.28 | 0.26 |

Based on the Voigt-Reuss-Hill (VRH) approximation[57], the bulk modulus, Young's modulus, shear modulus, and Poisson's ratio are derived, as shown in Table 5. The tobermorite 9 Å exhibits the highest overall elastic moduli, while the 14 Å shows the lowest values. This trend is mainly influenced by the interlayer spacing, the content of interlayer water molecules, and the strength of interlayer interactions. It indicates that as hydration decreases, the increase in interlayer distance leads to a decline in overall mechanical performance of tobermorite systems.

The radial distribution functions (RDFs) for Si-O, H-O, and Ca-O atomic pairs in tobermorite with 9 Å, 11 Å, and 14 Å are calculated by NEP and DFT. The RDF describes the probability of finding a specific type of particle at a given distance from a reference particle within a defined spatial region. In MD simulations, the RDF can be interpreted as the ratio of the local particle density in a spherical shell around a reference particle to the average bulk density. The position of the peaks in the RDF indicates the typical bond lengths between atomic pairs. The RDF is mathematically expressed as:

$$g(r) = \frac{\rho(r)}{\rho_0}, \#(9)$$

Here, $\rho(r)$ represents the local density of the target particles at a distance $r$ from the reference particle, and $\rho_0$ is the average density of the target particles in the system.

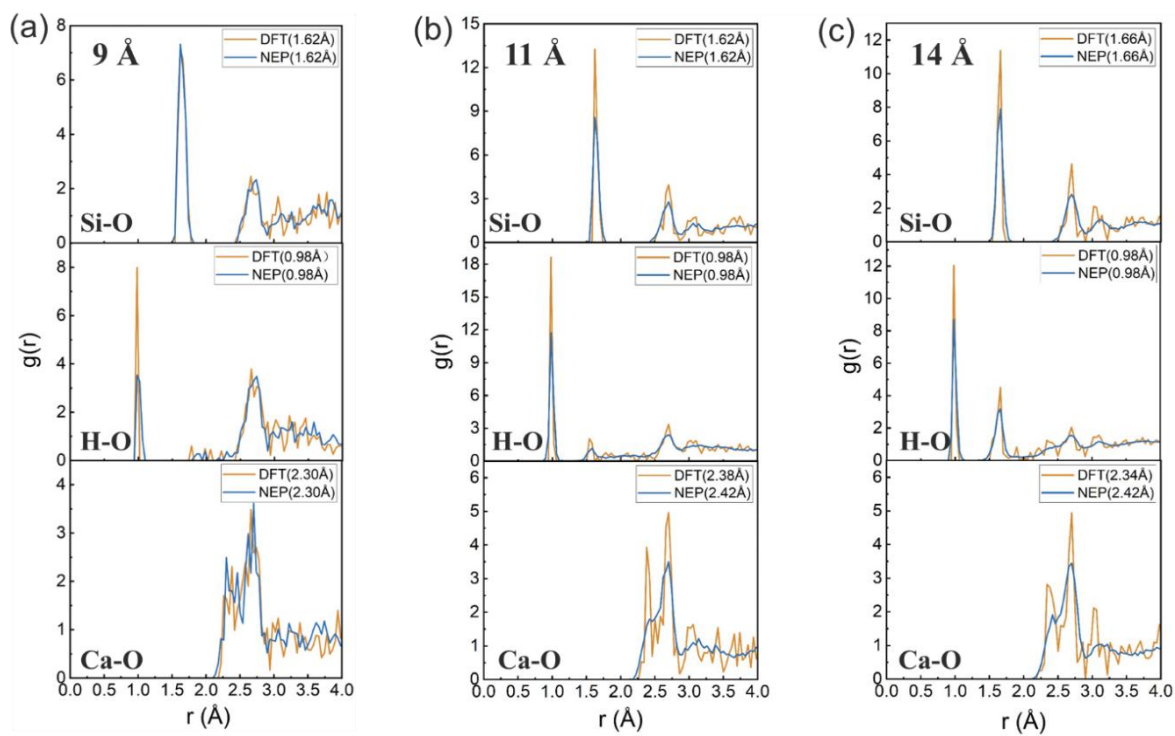

**Figure 7.** RDF of atomic pairs in tobermorite (a) 9 Å, (b) 11 Å, and (c) 14 Å, calculated by DFT and NEP (the values in the legends indicate the position of the first RDF peak).

As shown in the Figure 7, the NEP-predicted radial distribution functions exhibit good agreement with the DFT results. In particular, the positions of the first peaks for the Si-O and H-O bonds are nearly identical. For the Si-O pair in the 9 Å and 11 Å structures [Figure 7(a-b)], the first peak appears at ~1.62 Å, while in the 14 Å structure [Figure 7(c)], it shifts slightly to around 1.66 Å. Notably, for the H-O pair, a deviation is observed in the height of the first RDF peak between NEP and DFT, suggesting that the developed NEP model may have limitations in accurately describing the short-range potential energy surface related to hydrogen bonding. This discrepancy could stem from the insufficient representation of H-O short-range interactions by the basis functions and high-order angular descriptors used in NEP model. For the Ca-O pair, the first peak positions predicted by NEP for the 11 Å and 14 Å structures are slightly shifted to longer distances compared to DFT, but still lie within the range of 2.36-2.42 Å reported in previous simulations by Hou *et al* [58]. This shift may be attributed to the ionic nature of the Ca-O bond, where the high mobility of $Ca^{2+}$ ions leads to frequent bond breaking and reforming. Such interactions involve not only local bonding but also strong electrostatic and van der Waals interactions, which may not be fully captured by the NEP model. Overall, while NEP shows some deviations in the peak positions and heights for H-O and Ca-O bonds, it still reproduces the RDF profiles with reasonable accuracy and reliably captures the local bonding environments and coordination structures.

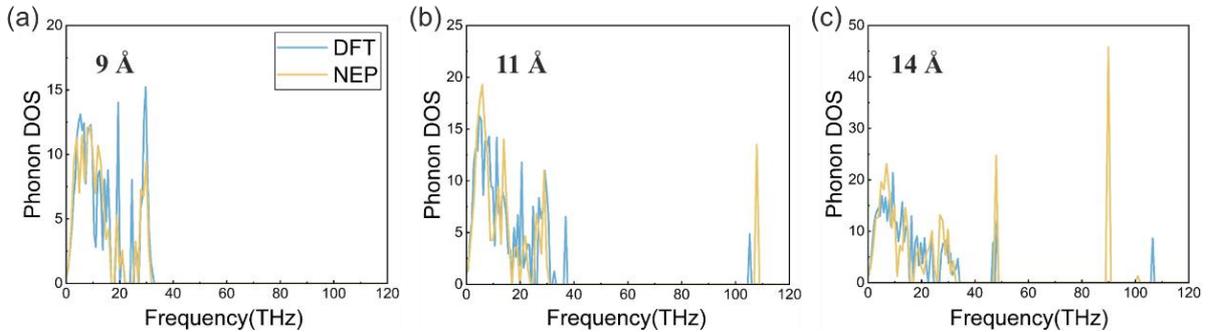

**Figure 8.** Phonon density of states (PDOS) of tobermorite for (a) 9 Å, (b) 11 Å, and (c) 14 Å structures, calculated using NEP and DFT.

The phonon density of states (PDOS) describes the distribution of phonons across different frequencies and is closely related to the thermodynamic properties of materials. In this study, the PDOS was computed using the force constant matrix obtained via the finite difference method, with an atomic displacement step size of 0.01 Å, as shown in Figure 8.

It can be seen that the PDOS for all three structures exhibits no imaginary frequencies,

indicating that the structures are stable under both DFT and NEP calculations. In the low-frequency region (< 40 THz), the NEP-predicted PDOS agrees well with DFT results, suggesting that the phonon vibration modes are rich. These low-frequency modes are likely associated with Si-O-Si bond vibrations, strong interlayer interactions, and Si-O-H bending vibrations, all of which involve angular displacements and typically manifest at lower frequency region. In the high-frequency region of the 11 Å and 14 Å structures [Figure 8(b) and (c)], both NEP and DFT results show distinct isolated peaks. These peaks are likely due to the stretching vibrations of the hydroxyl (H-O) bonds. Since hydroxyl bonds are lighter and stronger, their vibrations involve only direct changes in bond length, resulting in higher vibration frequencies. Moreover, H-O stretching vibrations typically are not strongly coupled with vibrations of heavier atoms like Si-O and Ca-O, which is why both NEP and DFT can capture this characteristic peak. Overall, the results demonstrate that the NEP model effectively captures both the long-wavelength phonon dispersion and the key vibrational characteristics across a wide frequency range, validating its reliability in describing the vibrational properties of tobermorite structures.

### 4.3 Application of NEP in MD

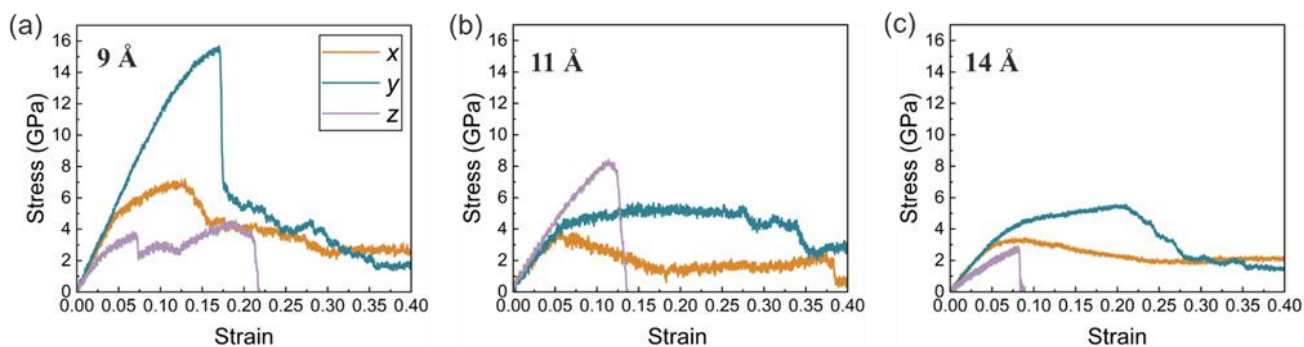

**Figure 9.** Tensile stress-strain curves of tobermorite (a) 9 Å, (b) 11 Å, and (c) 14 Å.

We then perform large-scale uniaxial tensile simulations along the *x*, *y*, and *z* directions on orthogonal tobermorite supercells using the developed NEP model. The tensile loading was applied at a constant velocity of $3.3 \times 10^{-5}$ Å/ps, and the corresponding stress-strain curves are shown in Figure 9. As the interlayer distance increases, the stress-strain responses along the *x*, *y*, and *z* directions exhibit different trends, highlighting the pronounced directional anisotropy in the mechanical properties of the layered tobermorite structure. For the tobermorite 9 Å structure [Figure 9(a)], the tensile strength in the *y* direction reaches approximately 16 GPa, significantly higher than in the *x* and *z* directions, indicating strong resistance to deformation along the interlayer bonding direction. However, the sharp stress drop following the peak stress suggests limited plasticity and a brittle failure mode. The pronounced differences in tensile response among the three directions further confirm the anisotropic mechanical behavior of the 9 Å structure. In the case of the tobermorite 11 Å [Figure 9(b)], the stress-strain curve in the *x* direction exhibits an initial linear increase up to 3.36 GPa at a strain of 0.05, followed by fluctuations and gradual stress decrease until structural failure occurs at a strain of ~ 0.18. Notably, the ultimate tensile strength of tobermorite 11 Å in the *z* direction exceeds that of both the 9 Å and 14 Å structures, consistent with the elastic constant trend $C_{33} > C_{22} > C_{11}$. This enhanced strength arises from the presence of robust Si-O-Si bonds in the 11 Å structure, which effectively redistribute stress during deformation along the *z* direction and delay the local structural failure. For the tobermorite 14 Å structure [Figure 9(c)], the peak stresses observed in all three directions are consistently lower than those in the 9 Å and 11 Å cases, indicating a degradation in mechanical performance with increasing interlayer spacing. This reduction in strength suggests a higher propensity for interlayer slip or failure, which aligns with the decrease in elastic modulus reported in Table 4.

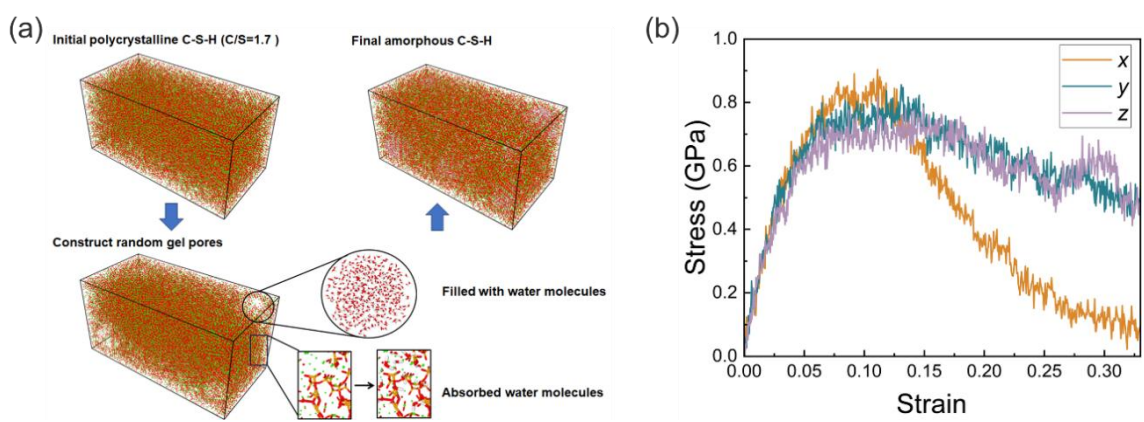

**Figure 10.** (a) C-S-H modeling process; (b) Stress-strain curves at a stretching velocity of $2 \times 10^{-4}$ Å/ps.

The NEP model was further extended to simulate C-S-H systems. As shown in Figure 10(a), a C-S-H model with a Ca/Si of 1.7 was constructed using our developed CSH modeling program[59]. A polycrystalline structure was then built by assembling 187 C-S-H unit cells within a simulation box of dimensions $20 \times 10 \times 10$ nm$^3$. To more realistically capture the porous nature of C-S-H, gel pores with diameters ranging from 1 to 4 nm were introduced into the polycrystalline matrix, and pre-constructed water spheres were embedded within these pores. Additionally, extra water molecules were added to the C-S-H matrix to more accurately represent the water distribution within gel pores and its influence on the C-S-H structure. The final polycrystalline model contained a total of 115,303 atoms.

Subsequently, large-scale uniaxial tensile simulations were carried out in the *x*, *y*, and *z* directions at a constant velocity of 2 × 10$^{-4}$ Å/ps, and the resulting stress-strain curves are shown in Figure 10(b). In the initial linear elastic regime, the stress-strain curves in all three directions exhibit significant overlap, indicating isotropic elastic behavior, which is consistent with the amorphous nature of the C-S-H model. The tensile elastic modulus estimated at 2% strain is approximately 17.85 GPa, which closely matches the experimental value of 17.35 GPa reported by Fu *et al* [17]. As the strain reaches around 0.1, the peak stress values in the three directions are about 0.9, 0.8, and 0.7 GPa, respectively, which are in good agreement with the simulation results by Zhou *et al*[60]. Beyond this point, the material enters the plastic deformation stage, and the stress-strain curves start to diverge due to the random distribution of C-S-H nanocrystals and defects within the system. With the trained NEP potential, we have preliminarily achieved large-scale uniaxial tensile simulations of amorphous C-S-H systems. The study found that the NEP model demonstrates good predictive accuracy in the small-strain elastic stage and near the stress peak around 0.1 strain, which potentially correlates with the volumetric deformation range (±10%) included in the training dataset. However, it should be noted that since the training data did not cover configurations under large plastic deformation, the predictive accuracy of the potential function significantly decreases once the strain exceeds 0.1. Therefore, simulation results in this stage should currently be regarded as qualitative references, and related conclusions must be interpreted with caution. Future work should expand the training dataset to include larger strain levels, defect generation, and structural failure processes, thereby enhancing the model's predictive capability in the large-strain regime.

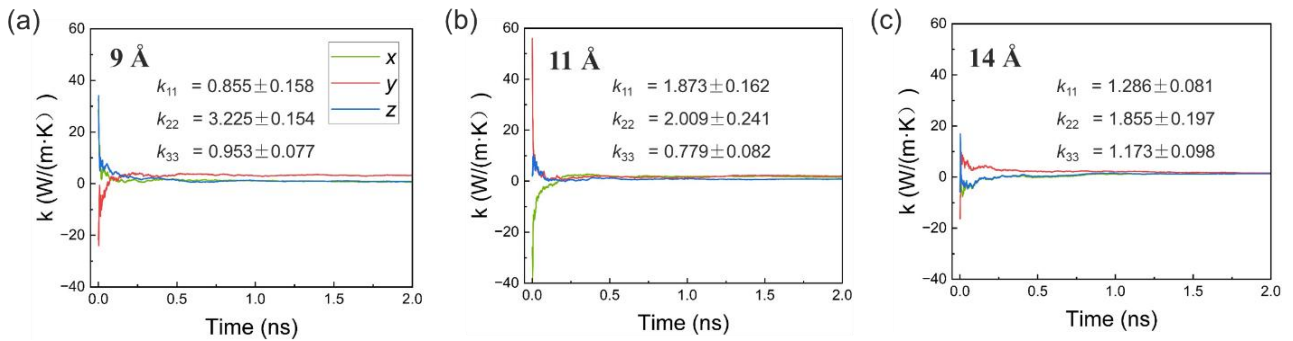

**Figure 11.** The thermal conductivity of tobermorite (a) 9 Å, (b) 11 Å, and (c) 14 Å, obtained from five independent simulations for each case.

The previous PDOS analysis demonstrates that the developed NEP model accurately captures phonon behavior in the low-frequency region, and low-frequency phonons are the main carriers of heat conduction. Thus, the NEP model provide a reliable foundation for the subsequent study of thermal transport properties. Building on this, we employed the trained NEP model to calculate the thermal conductivity of tobermorite using the homogeneous non-equilibrium molecular dynamics (HNEMD) method under conditions of 300 K and zero pressure. In this process, we considered the diffusion mechanism and constructed cubic cell models containing 7440, 8800, and 7800 atoms for the 9 Å, 11 Å, and 14 Å structures, respectively. Figure 11 presents the time evolution of thermal conductivity obtained from five independent simulations, where a driving force of $5\times10^{-5}$ Å$^{-1}$ was applied along different directions. It can be seen that the thermal conductivity fluctuates significantly at the initial stage and then gradually converges, showing anisotropy in different directions, and the numerical results are close to those reported in previous studies[24, 26]. For tobermorite with a 9 Å interlayer spacing [Figure 11(a)], the anisotropy in thermal conductivity is relatively pronounced, but the *y*-direction shows significantly higher conductivity than the *x* and *z* directions. This may be due to the smaller interlayer spacing of the 9 Å and the absence of interlayer water molecules, which enhance interlayer interactions and result in denser atomic packing, thereby reducing phonon scattering., making it easier for phonons to propagate along the *y* direction. For the 11 Å and 14 Å structures [Figure 11(b) and (c)], it can be observed that the thermal conductivity in the *y* direction is higher than in the *x* and *z* directions, indicating stronger thermal conductivity along this direction. Furthermore, in terms of the anisotropic trend, the relationship $k_{22} > k_{11} > k_{33}$ is consistent with the findings of Hong *et al* [26].This may be attributed to the predominant alignment of SiO$_4$ tetrahedra along the *y*-direction, where strong Si-O-Si covalent bonds enhance intralayer interactions and thus facilitate more efficient phonon transport. On the other hand, the larger interlayer distance in the *z-direction*, coupled with weak interactions from calcium ions and water molecules between the layers, limits the transfer of thermal vibrations, resulting in lower thermal conductivity in the *z-direction*.

## 5. Conclusions

We have developed an effective machine learning potential model for tobermorite and C-S-H systems based on NEP framework, incorporating an active learning strategy. The model was rigorously validated through a series of physical property evaluations and large-scale simulation tests.

Based on the comprehensive results, the following key conclusions can be drawn:

(1) NEP significantly reduces computational resource requirements by achieving DFT-level accuracy with only a few hundred training data points.

(2) GPU-accelerated NEP simulations exhibit exceptional computational efficiency, markedly outperforming existing MLP frameworks such as DP, thus enabling realistic, large-scale molecular dynamics studies.

(3) The model accurately predicts structural, mechanical, and thermal properties, including equations of state, elastic constants, radial distribution functions, phonon density of states, and thermal conductivity.

(4) Initial applications of NEP to tensile simulations of amorphous C-S-H systems demonstrate its versatility and potential for extensive studies of cementitious materials under practical, large-scale conditions.

Overall, the NEP model significantly lowers the computational cost of neural network training while satisfactorily balancing accuracy and efficiency. It enables efficient simulations of hundreds of thousands of atoms using only a single GPU card. Although the current study focuses mainly on tobermorite minerals, the NEP can be extended to other cementitious materials by including more types of C-S-H defect structures in the training process. These results provide a powerful tool for accurate and large-scale MD simulations of cement-based materials.

## Acknowledgement

The authors acknowledge financial supports from Natural Science Foundation of Shandong Province, China (Grant No.: ZR2021MA067) and Shandong Agricultural University Young Faculty Development Plan. We also thank Supercomputing Center in Shandong Agricultural University for technical support.

## References

[1] Q Luo, Y Xiang, Q Yang, T Liang, Y Xie. Molecular simulation of calcium-silicate-hydrate and its applications: A comprehensive review, Construction and Building Materials. 409 (2023) 134137.

[2] R K Mishra, A K Mohamed, D Geissbühler, H Manzano, T Jamil, R Shahsavari, A G Kalinichev, S Galmarini, L Tao, H Heinz, R Pellenq, A C T van Duin, S C Parker, R J Flatt, P Bowen. cemff: A force field database for cementitious materials including validations, applications and opportunities, Cement and Concrete Research. 102 (2017) 68-89.

[3] X M Aretxabaleta, J López-Zorrilla, I Etxebarria, H Manzano. Multi-step nucleation pathway of

C-S-H during cement hydration from atomistic simulations, Nature Communications. 14 (2023) 7979.

[4] M J Abdolhosseini Qomi, F-J Ulm, R J M Pellenq. Physical Origins of Thermal Properties of Cement Paste, Physical Review Applied. 3 (2015) 064010.

[5] M J Abdolhosseini Qomi, L Brochard, T Honorio, I Maruyama, M Vandamme. Advances in atomistic modeling and understanding of drying shrinkage in cementitious materials, Cement and Concrete Research. 148 (2021) 106536.

[6] J Xu, X Chen, G Yang, X Niu, F Chang, G Lacidogna. Review of research on micromechanical properties of cement-based materials based on molecular dynamics simulation, Construction and Building Materials. 312 (2021) 125389.

[7] Y Yan, G Geng. Does nano basic building-block of C-S-H exist? – A review of direct morphological observations, Materials & Design. 238 (2024) 112699.

[8] G Renaudin, J Russias, F Leroux, F Frizon, C Cau-dit-Coumes. Structural characterization of C-S-H and C-A-S-H samples—Part I: Long-range order investigated by Rietveld analyses, Journal of Solid State Chemistry. 182 (2009) 3312-3319.

[9] C Rößler, F Steiniger, H M Ludwig. Characterization of C-S-H and C-A-S-H phases by electron microscopy imaging, diffraction, and energy dispersive X‐ray spectroscopy, Journal of the American Ceramic Society. 100 (2017) 1733-1742.

[10] J Li, W Zhang, K Garbev, G Beuchle, P J M Monteiro. Influences of cross-linking and Al incorporation on the intrinsic mechanical properties of tobermorite, Cement and Concrete Research. 136 (2020) 106170.

[11] M Izadifar, F Königer, A Gerdes, C Wöll, P Thissen. Correlation between Composition and Mechanical Properties of Calcium Silicate Hydrates Identified by Infrared Spectroscopy and Density Functional Theory, The Journal of Physical Chemistry C. 123 (2019) 10868-10873.

[12] D Tunega, A Zaoui. Understanding of bonding and mechanical characteristics of cementitious mineral tobermorite from first principles, J Comput Chem. 32 (2011) 306-314.

[13] S Hajilar, B Shafei. Nano-scale investigation of elastic properties of hydrated cement paste constituents using molecular dynamics simulations, Computational Materials Science. 101 (2015) 216-226.

[14] X Wang, T Li, W Xie, L Zhang, D Li, F Xing. Molecular dynamics study on the structure and mechanical properties of tobermorite, Materials Science and Engineering: B. 299 (2024) 116930.

[15] R Shahsavari, M J Buehler, R J M Pellenq, F J Ulm. First‐Principles Study of Elastic Constants and Interlayer Interactions of Complex Hydrated Oxides: Case Study of Tobermorite and Jennite, Journal of the American Ceramic Society. 92 (2009) 2323-2330.

[16] Q Zheng, J Jiang, J Yu, X Li, S Li. Aluminum-Induced Interfacial Strengthening in Calcium Silicate Hydrates: Structure, Bonding, and Mechanical Properties, ACS Sustainable Chemistry & Engineering. 8 (2020) 2622-2631.

[17] J Fu, S Kamali-Bernard, F Bernard, M Cornen. Comparison of mechanical properties of C-S-H and portlandite between nano-indentation experiments and a modeling approach using various simulation techniques, Composites Part B: Engineering. 151 (2018) 127-138.

[18] Y Li, H Pan, Z Li. Ab initio metadynamics simulations on the formation of calcium silicate aqua complexes prior to the nuleation of calcium silicate hydrate, Cement and Concrete Research.

156 (2022) 106767.

[19] I-H Svenum, I G Ringdalen, F L Bleken, J Friis, D Höche, O Swang. Structure, hydration, and chloride ingress in C-S-H: Insight from DFT calculations, Cement and Concrete Research. 129 (2020) 105965.

[20] J Behler. Perspective: Machine learning potentials for atomistic simulations, J Chem Phys. 145 (2016) 170901.

[21] R T Cygan, J-J Liang, A G Kalinichev. Molecular Models of Hydroxide, Oxyhydroxide, and Clay Phases and the Development of a General Force Field, J Phys Chem. 108 (2004) 1255-1266.

[22] R Shahsavari, R J Pellenq, F J Ulm. Empirical force fields for complex hydrated calcio-silicate layered materials, Phys Chem Chem Phys. 13 (2011) 1002-1011.

[23] D Fan, S Yang. Mechanical properties of C-S-H globules and interfaces by molecular dynamics simulation, Construction and Building Materials. 176 (2018) 573-582.

[24] Y Yang, Y Wang, J Cao. Prediction and evaluation of thermal conductivity in nanomaterial-reinforced cementitious composites, Cement and Concrete Research. 172 (2023) 107240.

[25] S-N Hong, C-J Yu, U-S Hwang, C-H Kim, B-H Ri. Effect of porosity and temperature on thermal conductivity of jennite: A molecular dynamics study, Materials Chemistry and Physics. 250 (2020) 123146.

[26] S-N Hong, C-J Yu, K-C Ri, J-M Han, B-H Ri. Molecular dynamics study of the effect of moisture and porosity on thermal conductivity of tobermorite 14 Å, International Journal of Thermal Sciences. 159 (2021) 106537.

[27] R K Mishra, A K Mohamed, D Geissbühler, H Manzano, T Jamil, R Shahsavari, A G Kalinichev, S Galmarini, L Tao, H Heinz. cemff: A force field database for cementitious materials including validations, applications and opportunities, Cement and Concrete Research. 102 (2017) 68-89.

[28] M Valavi, Z Casar, A Kunhi Mohamed, P Bowen, S Galmarini. Molecular dynamic simulations of cementitious systems using a newly developed force field suite ERICA FF, Cement and Concrete Research. 154 (2022) 106712.

[29] E Duque-Redondo, P A Bonnaud, H Manzano. A comprehensive review of C-S-H empirical and computational models, their applications, and practical aspects, Cement and Concrete Research. 156 (2022) 106784.

[30] Z Qi, X Sun, Z Sun, Q Wang, D Zhang, K Liang, R Li, D Zou, L Li, G Wu, W Shen, S Liu. Interfacial Optimization for AlN/Diamond Heterostructures via Machine Learning Potential Molecular Dynamics Investigation of the Mechanical Properties, ACS Appl Mater Interfaces. 16 (2024) 27998-28007.

[31] K Zhu, Z Zhang. Equivariance is essential, local representation is a need: A comprehensive and critical study of machine learning potentials for tobermorite phases, Computational Materials Science. 246 (2025) 113363.

[32] K Kobayashi, H Nakamura, A Yamaguchi, M Itakura, M Machida, M Okumura. Machine learning potentials for tobermorite minerals, Computational Materials Science. 188 (2021) 110173.

[33] Y Zhou, H Zheng, W Li, T Ma, C Miao. A deep learning potential applied in tobermorite phases and extended to calcium silicate hydrates, Cement and Concrete Research. 152 (2022) 106685.

[34] W Li, Y Zhou, L Ding, P Lv, Y Su, R Wang, C Miao. A deep learning-based potential developed for calcium silicate hydrates with both high accuracy and efficiency, Journal of Sustainable Cement-Based Materials. 12 (2023) 1335-1346.

[35] W Li, C Xiong, Y Zhou, W Chen, Y Zheng, W Lin, J Xing. Insights on the mechanical properties and failure mechanisms of calcium silicate hydrates based on deep-learning potential molecular dynamics, Cement and Concrete Research. 186 (2024) 107690.

[36] H Wang, L Zhang, J Han, W E. DeePMD-kit: A deep learning package for many-body potential energy representation and molecular dynamics, Computer Physics Communications. 228 (2018) 178-184.

[37] L Zhang, J Han, H Wang, R Car, W E. Deep Potential Molecular Dynamics: A Scalable Model with the Accuracy of Quantum Mechanics, Phys Rev Lett. 120 (2018) 143001.

[38] K Zhu. Performance Comparisons of NequIP and DPMD Machine Learning Interatomic Potentials for Tobermorites, Computational Materials Science. 244 (2024) 113212.

[39] S Batzner, A Musaelian, L Sun, M Geiger, J P Mailoa, M Kornbluth, N Molinari, T E Smidt, B Kozinsky. E(3)-equivariant graph neural networks for data-efficient and accurate interatomic potentials, Nat Commun. 13 (2022) 2453.

[40] Z Fan, Y Wang, P Ying, K Song, J Wang, Y Wang, Z Zeng, K Xu, E Lindgren, J M Rahm. GPUMD: A package for constructing accurate machine-learned potentials and performing highly efficient atomistic simulations, The Journal of Chemical Physics. 157 (2022) 114801.

[41] Z Fan, Z Zeng, C Zhang, Y Wang, K Song, H Dong, Y Chen, T Ala-Nissila. Neuroevolution machine learning potentials: Combining high accuracy and low cost in atomistic simulations and application to heat transport, Physical Review B. 104 (2021) 104309.

[42] P Ying, C Qian, R Zhao, Y Wang, F Ding, S Chen, Z Fan. Advances in modeling complex materials: The rise of neuroevolution potentials, arXiv preprint arXiv:250111191. (2025).

[43] Y Wang, Z Fan, P Qian, M A Caro, T Ala-Nissila. Quantum-corrected thickness-dependent thermal conductivity in amorphous silicon predicted by machine learning molecular dynamics simulations, Physical Review B. 107 (2023) 054303.

[44] Z Li, J Wang, H Dong, Y Zhou, L Liu, J-Y Yang. Mechanistic insights into water filling effects on thermal transport of carbon nanotubes from machine learning molecular dynamics, International Journal of Heat and Mass Transfer. 235 (2024) 126152.

[45] Z Qi, X Sun, Z Sun, Q Wang, D Zhang, K Liang, R Li, D Zou, L Li, G Wu. Interfacial optimization for AlN/diamond heterostructures via machine learning potential molecular dynamics investigation of the mechanical properties, ACS Appl Mater Interfaces. 16 (2024) 27998-28007.

[46] K Xu, Y Hao, T Liang, P Ying, J Xu, J Wu, Z Fan. Accurate prediction of heat conductivity of water by a neuroevolution potential, The Journal of Chemical Physics. 158 (2023).

[47] Z Zhao, M Yi, W Guo, Z Zhang. General-purpose neural network potential for Ti-Al-Nb alloys towards large-scale molecular dynamics with ab initio accuracy, Physical Review B. 110 (2024) 184115.

[48] R Zhao, S Wang, Z Kong, Y Xu, K Fu, P Peng, C Wu. Development of a neuroevolution machine learning potential of Pd-Cu-Ni-P alloys, Materials & Design. 231 (2023) 112012.

[49] V Wang, N Xu, J-C Liu, G Tang, W-T Geng. VASPKIT: A user-friendly interface facilitating

high-throughput computing and analysis using VASP code, Computer Physics Communications. 267 (2021) 108033.

[50] A H Larsen, J J Mortensen, J Blomqvist, I E Castelli, R Christensen, M Dułak, J Friis, M N Groves, B Hammer, C Hargus. The atomic simulation environment-a Python library for working with atoms, Journal of Physics: Condensed Matter. 29 (2017) 273002.

[51] J Ziegler. JP Biersack and U. Littmark, The stopping and range of ions in solids. 1 (1985).

[52] https://github.com/HYinSD/CSH-NEP.

[53] M Greenacre, P J Groenen, T Hastie, A I d'Enza, A Markos, E Tuzhilina. Principal component analysis, Nature Reviews Methods Primers. 2 (2022) 100.

[54] E Bonaccorsi, S Merlino, A R Kampf. The Crystal Structure of Tobermorite 14 Å (Plombierite), a C-S-H Phase, Journal of the American Ceramic Society. 88 (2005) 505-512.

[55] S Merlino, E Bonaccorsi, T Armbruster. The real structures of clinotobermorite and tobermorite 9 Å: OD character, polytypes, and structural relationships, European Journal of Mineralogy. 12 (2000) 411-429.

[56] S Merlino, E Bonaccorsi, T Armbruster. The real structure of tobermorite 11A: normal and anomalous forms, OD character and polytypic modifications, European Journal of Mineralogy. 13 (2001) 577-590.

[57] H Manzano, J S Dolado, A Ayuela. Elastic properties of the main species present in Portland cement pastes, Acta Materialia. 57 (2009) 1666-1674.

[58] D Hou, T Zhao, Z Jin, H Ma, Z Li, L Q Chen. Molecular Simulation of Calcium Silicate Composites: Structure, Dynamics, and Mechanical Properties, Journal of the American Ceramic Society. 98 (2014) 758-769.

[59] S Wang, F Ren, Y Song, G Papadakis, Y Yang, H Yin. Issues of standardizing C-S-H molecular models: Random defect distribution and its effects on material performance, Construction and Building Materials. 470 (2025) 140527.

[60] A Zhou, J Kang, R Qin, H Hao, T Liu, Z Yu. Weaving the next-level structure of calcium silicate hydrate at the submicron scale via a remapping algorithm from coarse-grained to all-atom model, Cement and Concrete Research. 180 (2024) 107501.